\documentclass[graybox]{svmult}

\usepackage{mathptmx}       
\usepackage{helvet}         
\usepackage{courier}        
\usepackage{type1cm}        
\usepackage{makeidx}         
\usepackage{graphicx}        
\usepackage{multicol}        
\usepackage[bottom]{footmisc}
\usepackage{epsfig}

\makeindex             

\usepackage{verbatim}
\newcommand{\gev}{{\rm \,Ge\kern-0.125em V}}
\newcommand{\mpl}{M_{\rm {Pl}}}

\begin{document}

\title*{Inflation after Planck and BICEP2}
\author{Raghavan Rangarajan}
\institute{Raghavan Rangarajan \at Theoretical Physics Division, Physical Research Laboratory, Navrangpura, Ahmedabad, \email{raghavan@prl.res.in}
}
%
%
\maketitle

\abstract{
We discuss the inflationary paradigm, how it can be tested, and how various models of inflation fare in the light of data from Planck and BICEP2.
We introduce inflation and reheating, and discuss temperature and
polarisation anisotropies in the cosmic microwave background radiation due
to quantum fluctuations during inflation.  Fitting observations of the
anisotropies with theoretical realisations
obtained by varying various parameters of the curvature power spectrum and cosmological parameters enables one to obtain the allowed ranges
of these parameters.  We  
discuss how to relate these parameters to inflation models which allows one to rule in or out specific models of
inflation.  
}

\section{Introduction}
\label{sec:1}

\begin{figure}
\sidecaption
\includegraphics[width=7cm]{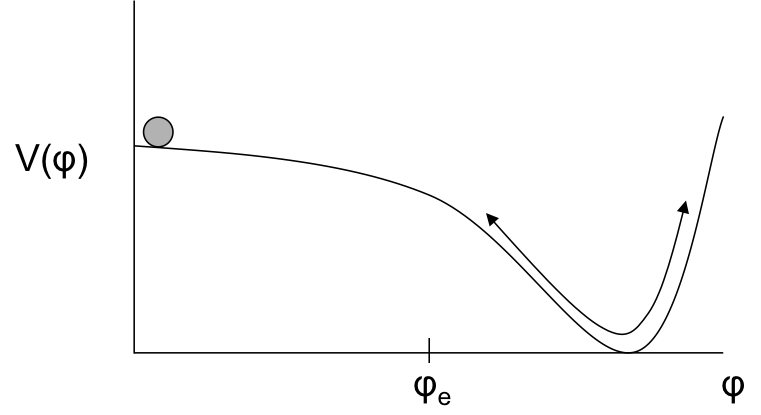}
%
%
%
\caption{A cartoon of the inflaton potential.  For $\varphi<\varphi_e$ the potential is flat and the field rolls slowly.  For $\varphi>\varphi_e$
the field oscillates in its potential and decays.}
\label{fig:inflaton_potential}
\end{figure}

Inflation is a period of accelerated expansion in the early Universe when the energy density of the Universe is dominated by the potential energy of a slowly moving scalar field (or fields) called the inflaton \cite{kolbturner,baumann}.  It leads to an approximately exponential expansion of the Universe.  Inflation happened when
the energy density in the Universe, which is approximately the potential energy of the inflaton field
$\varphi$, was of the order the GUT scale $[\sim (10^{16} \gev)^4]$ or less, i.e. when the age of the Universe was $10^{-38}$ s or more.  In Fig. \ref{fig:inflaton_potential} we show a cartoon of the potential of the field $\varphi$.
When $\varphi<\varphi_e$ the field is rolling slowly and the Universe is in the inflationary era.  

During inflation, the scale factor $R(t)$ of the Universe, 
which indicates 
how distances are increasing in an expanding Universe, varies as $\sim\exp(H_I t)$ where $H_I$ is the approximately constant
Hubble parameter during the inflationary era.  For GUT scale inflation, inflation lasts for at least 60 e-foldings, i.e. the duration is
at least $60 H_I^{-1}$.  During the inflationary era the number density 
of all species goes to 0 because of the exponential increase in the volume.
After $\varphi$ crosses $\varphi_e$ the field oscillates in its potential, it decays thereby repopulating the
Universe with particles, and then the decay products thermalise.  This is referred to as the reheating era, and
the Universe becomes radiation dominated thereafter.  The potential in Fig. \ref{fig:inflaton_potential} describes a new inflation or hilltop
inflation model.  There are many different models of inflation with different forms of the inflaton potential \cite{martin_encycl}.

Inflation was proposed to solve the horizon problem and the flatness problem in cosmology.  The horizon problem is the following:
why is the cosmic microwave background radiation (CMBR) almost uniform at separation angles greater
than $1^\circ$ in the sky, when the corresponding regions would not normally have been in causal contact  when the CMBR was `formed'.  Inflation expands a causally connected
region exponentially to a very large volume at a very early time and the CMBR we receive today from different directions 
comes to us  from regions within this causally connected expanded volume and hence can be correlated.  
The flatness problem is 
that for the spatial curvature of the Universe to be as close to 0 as it is today
it needs to be fine-tuned to be extremely small at the Planck time,
since it diverges from 0 as the Universe evolves.  Inflation drives the spatial curvature to be extremely small by the end of inflation so
that one does not need to invoke any fine-tuning in the early Universe to explain why it is so small today.  
\footnote{
Note that it has been argued in Ref. \cite{Carroll2014} that 
using
the Gibbons-Hawking-Stewart measure 
for trajectories of universes
implies that flat universes are generic, and
so the curvature term $\Omega_k\approx 0$ is very probable.  Also see Ref. \cite{SchiffrinWald} for a 
criticism of the use of this measure in cosmology.}
Besides solving these problems inflation provides, as a bonus, a mechanism for generating the initial density perturbations needed to seed the large scale structure in the Universe today.

\section{Observational Consequences of Inflation}
\label{sec:2}

During inflation there are quantum fluctuations in the inflaton field and in the spacetime metric.  
The former and scalar
perturbations of the metric give rise to perturbations on various length scales 
in the energy density of matter (non-relativistic
particles) and radiation (relativistic particles) after inflation.  Tensor perturbations of the metric are
associated with gravitational waves.  Perturbations in the matter energy density grow with time and give
rise to the galaxies, clusters, and other forms of large scale structure.  Perturbations in the radiation and 
matter energy density, and gravitational waves, are responsible for the anisotropy 
of the CMBR.  Thus
observations of large scale structure and the anisotropy of the CMBR allow us to test the inflationary
paradigm and models \cite{DodelsonBook}.

\subsection{Perturbations in the Matter Energy Density and Large Scale Structure}

A spatial perturbation in the matter energy density can be expressed as
\begin{equation}
\delta({\bf x})\equiv\frac{\delta\rho}{\rho}({\bf x}) =\frac{\rho({\bf x})-\rho_0}{\rho_0}
\end{equation}
where $\rho_0$ is the mean energy density.  If $\delta({\bf k})$ is the Fourier transform of $\delta({\bf x}) $
then the matter power spectrum 
\begin{equation}
P_\delta(k)\sim {\rm FT}\left[\langle\frac{\delta\rho}{\rho}({\bf x_1})
\frac{\delta\rho}{\rho}({\bf x_2})\rangle\right]\sim |\delta({\bf k})|^2\,,
\end{equation}
and is a measure of the perturbation on a scale 
$k^{-1}=|{\bf x_1}-{\bf x_2}|$.

Fluctuations in the inflaton field and scalar fluctuations of the metric are combined in a gauge invariant
way to obtain the scalar power spectrum, or curvature power spectrum.  
The matter power spectrum is a related quantity.  The
scalar power spectrum is given by
\begin{equation}
P(k)\sim\frac{H_I^4}{\mpl^2 \dot H_I}=A k^{n-1}\,,
\label{powerspectrum}
\end{equation}
where the slow rolling of the inflaton field implies that $n\sim 1$.  
($n=1$ is referred to as the Harrison-Zeldovich or scale invariant spectrum.)
While $n$ is often treated as approximately constant (independent of $k$),
more generally, inflation gives
\begin{equation}
n(k)=n(k_0) +\frac{1}{2} \frac{dn}{d(\ln k)}\biggr{|}_{k_0} \ln\left( \frac{k}{k_0}\right)\,.
\end{equation}
The variation of $n$ with $k$ is referred to as running of the spectral index.  One can 
compare the large scale
structure data with  
numerical simulations of structure formation using $P(k)$ as the initial perturbation to test
inflation and particular models of inflation.


\subsection{Fluctuations in the CMBR temperature}

\begin{figure}
\sidecaption
\includegraphics[width=7cm]{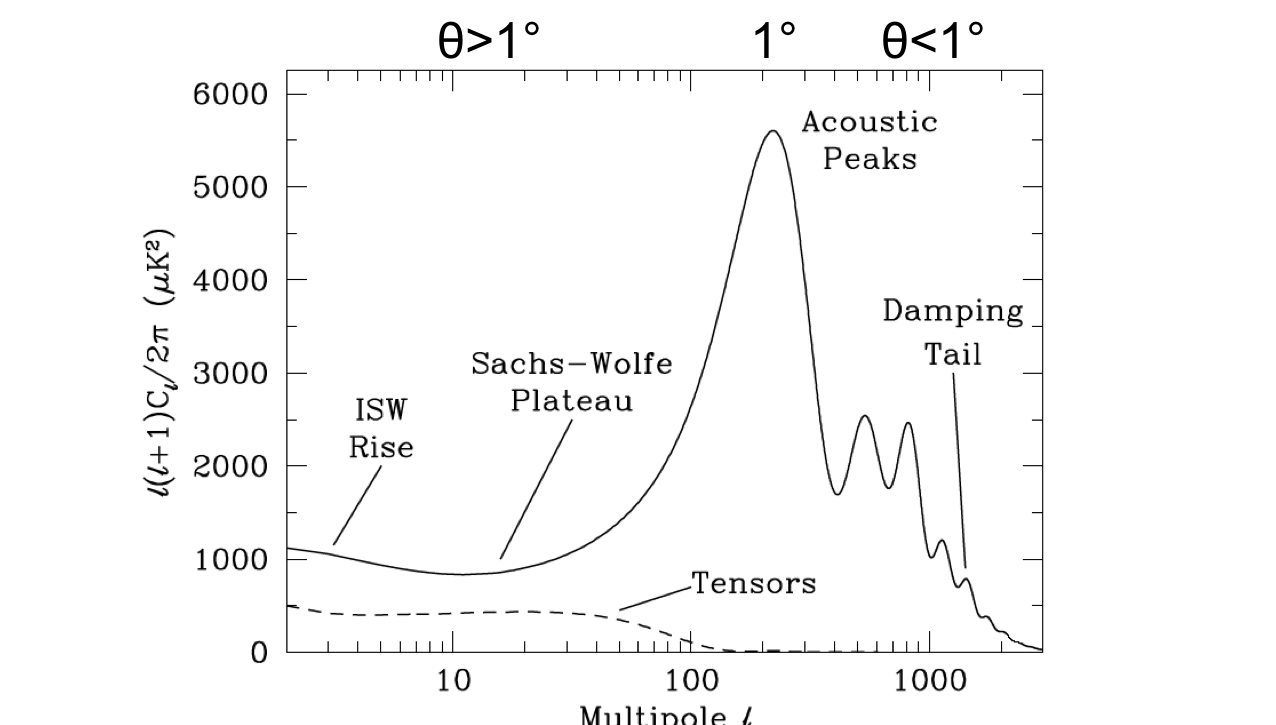}
%
%
%
\caption{
A theoretical CMBR angular power spectrum, $C_l$ for 
a standard $\Lambda$CDM model (using CAMB \cite{camb}). The plot shows 
the ISW rise,
Sachs-Wolfe plateau, acoustic peaks, and damping tail. The latter two are associated with small angle
anisotropies.  The contribution of
the tensor perturbations of the metric (gravitational waves), with an arbitrary normalisation, is also
shown. [From Ref. \cite{ScottSmootPDG2014}.]
}
\label{TheoreticalClPlot}
\end{figure}

The CMBR consists of photons that have been free-streaming through the Universe largely unscattered since the time of decoupling of photons and matter at $t_{\rm dec}\sim 380,000$ yr when neutral atoms formed. 
The temperature anisotropy in the CMBR 
is primarily
due to perturbations in 
the energy density of photons and non-relativistic matter at $t_{\rm dec}$, and in matter density perturbations
since $t_{\rm dec}$.  
These are referred to as the Sachs-Wolfe effect, and the Integrated Sachs-Wolfe (ISW) 
effect
respectively.  Furthermore, gravitational waves also contribute to the anisotropy in the temperature as well as the polarisation of the
CMBR.  If $T(\hat n_1)$ and $T(\hat n_2)$ are the CMBR temperature in the sky in directions
$\hat n_1$ and $\hat n_2$ separated by an angle $\theta$, and $T_0\sim 2.725$K is the mean temperature,
then
\begin{equation}
\frac{\langle T(\hat n_1) T(\hat n_2)\rangle_\theta - T_0^2}{T_0^2}
\sim
\sum_{l=2}^{\infty} (2l+1)C_l P_l(\cos\theta)
\end{equation}
where 
$\langle ... \rangle_\theta$ refers to an average across the sky of all pairs of directions separated by
an angle $\theta$, and
$C_l$ is called the angular power spectrum. 
Just as a power spectrum $P(k)$ is a measure of fluctuations on a scale $\lambda=k^{-1}$, $C_l$
is a measure of temperature fluctuations on an angular scale $\theta=180^\circ/l$.  $P_l$ is the Legendre polynomial.  On angles larger than $1^\circ$
the Sachs-Wolfe effect gives
\begin{equation}
C_l^{SW}\propto
\frac{\Omega_{\rm m}^2 H_0^2}{D_1} \int dk j_l^2[k(\eta_0-\eta_{\rm dec})] P(k)\,
\end{equation}
where $\Omega_{\rm m}$ is the ratio of the matter density today to the critical energy density, 
$H_0$ is the Hubble parameter today, $D_1$ is a growth factor, $j_l$ is the spherical Bessel 
function of order $l$, and $\eta_{0,{\rm dec}}$ are the conformal time today and at decoupling respectively \cite{DodelsonBook}.
Given the inflationary prediction of $P(k)$ as in Eq. (\ref{powerspectrum}) one can obtain the corresponding
$C_l$'s and compare these with observations.  
As an example, Fig. \ref{TheoreticalClPlot} shows a
theoretical angular power spectrum
using a standard $\Lambda$CDM model with 6 parameters. 

In addition to 2-point correlations of the CMBR temperature, self interactions of the scalar field
and non-linearities of General Relativity can give rise to non-trivial $n-$point correlations, which
are referred to as non-Gaussian fluctuations.   (For Gaussian fluctuations, all odd-point correlations are 0,
while even-point correlations are products of 2-point functions, as in Wick's theorem.  A weakly coupled inflaton
field is practically a free field and so inflation predicts largely Gaussian fluctuations of the inflaton and hence
of the temperature.) Three point correlations of the temperature are called
the bispectrum as they can be expressed as a product of two power spectra with a proportionality factor
$f_{\rm NL}$ called the non-Gaussianity parameter.  
\begin{equation}
\langle T(\hat n_1) T(\hat n_2) T(\hat n_3) \rangle \sim f_{\rm NL} \langle T(\hat n_i) T(\hat n_j)\rangle
\langle T(\hat n_k) T(\hat n_p)\rangle
\end{equation}
The above relation is symbolic -- 
more precisely the correlations are in  a parameter $\zeta(k)$, a gauge invariant quantity which is a combination of energy density and scalar metric fluctuations. 
For simple single field inflation models non-Gaussianity is small and   $f_{\rm NL} \ll1$
\cite{falketal,mald}.  In multifield inflation models and in models with non-standard kinetic
energy terms (k-inflation), one can have larger non-Gaussianities with $f_{\rm NL} >1$ (though the skewness 
$\sim$ [3 point correlation]/[2 point correlation]$^{3/2}$ is still small).  

The gravitational wave background due to inflation contributes to the angular power spectrum at large 
angles as shown in Fig. \ref{TheoreticalClPlot}.  Gravitational waves stretch and squeeze space which  
affects the wavelength of radiation and this produces a quadrupole anisotropy in the temperature distribution
of photons which are Thomson scattering off the electrons at the time of the last scattering
at $t_{\rm dec}$.  This ultimately gives rise to a net polarisation of the CMBR and to corresponding angular power spectra,
namely,
$C_l^{EE}, C_l^{BB}$ and $C_l^{TE}$ where $E$ and $B$ refer to divergence-free and curl-free polarisation components or modes, and $T$ refers to the temperature \cite{WayneHu,DodelsonBook}.
The power spectrum associated with gravitational waves is given by
\begin{equation}
P_T(k)\sim H_I^2/\mpl^2 = A_T k^{n_T}\,.
\end{equation}
The tensor-to-scalar ratio $r$ is defined as
\begin{equation}
r=\frac{P_T}{P}
\end{equation}
at some scale $k_*$.
Evidence for the effect of gravitational waves on the CMBR on scales larger than the causal horizon at decoupling
is treated as a smoking gun for inflation -- this is a prediction of inflation and
there is no other known mechanism for generating superhorizon gravitational waves.  
Currently we have not detected the gravitational wave background
and only have an upper bound on $r$.  A measurement of $r$ is also important because it
implies the energy scale of inflation since this is given by
\begin{equation}
V_I^\frac{1}{4} =\left(\frac{r}{0.12}\right)^\frac14 1.9 \times 10^{16} \gev\,.
\end{equation}


\section{CMBR Observations and Inflation}
\label{section3}

There were many experiments in the 70s and 80s to detect the CMBR temperature
anisotropy but they could not detect it.
There was concern then that if the anisotropy was not detected or was very small, then our understanding of 
structure formation would need to be revised as the 
promordial density perturbation which is the seed for structure formation is expected to have the 
same amplitude as
the large angle temperature
anisotropy.  COBE was the second satellite based experiment to study the cosmic microwave background radiation 
(Relikt 1 of the Soviet Union was the first)  and in 1992 it detected variations in the CMBR temperature across the sky at a level $\sim10^{-5}$ \cite{COBE1,COBE4} consistent with theories of structure formation. (A re-analysis of the data of
Relikt 1 also gave a detection of a similar anisotropy \cite{RELIKT}.)  There were subsequently many other experiments.
For example, MAT, BOOMERanG and MAXIMA detected the first peak in the angular power spectrum in 1999-2000, while 
DASI and CBI detected the $E$ mode polarisation of the CMBR in 2002 and 2004.  More recent experiments are
WMAP, Planck and BICEP which we shall discuss in greater detail below.

\begin{figure}
\sidecaption
\includegraphics[width=7cm]{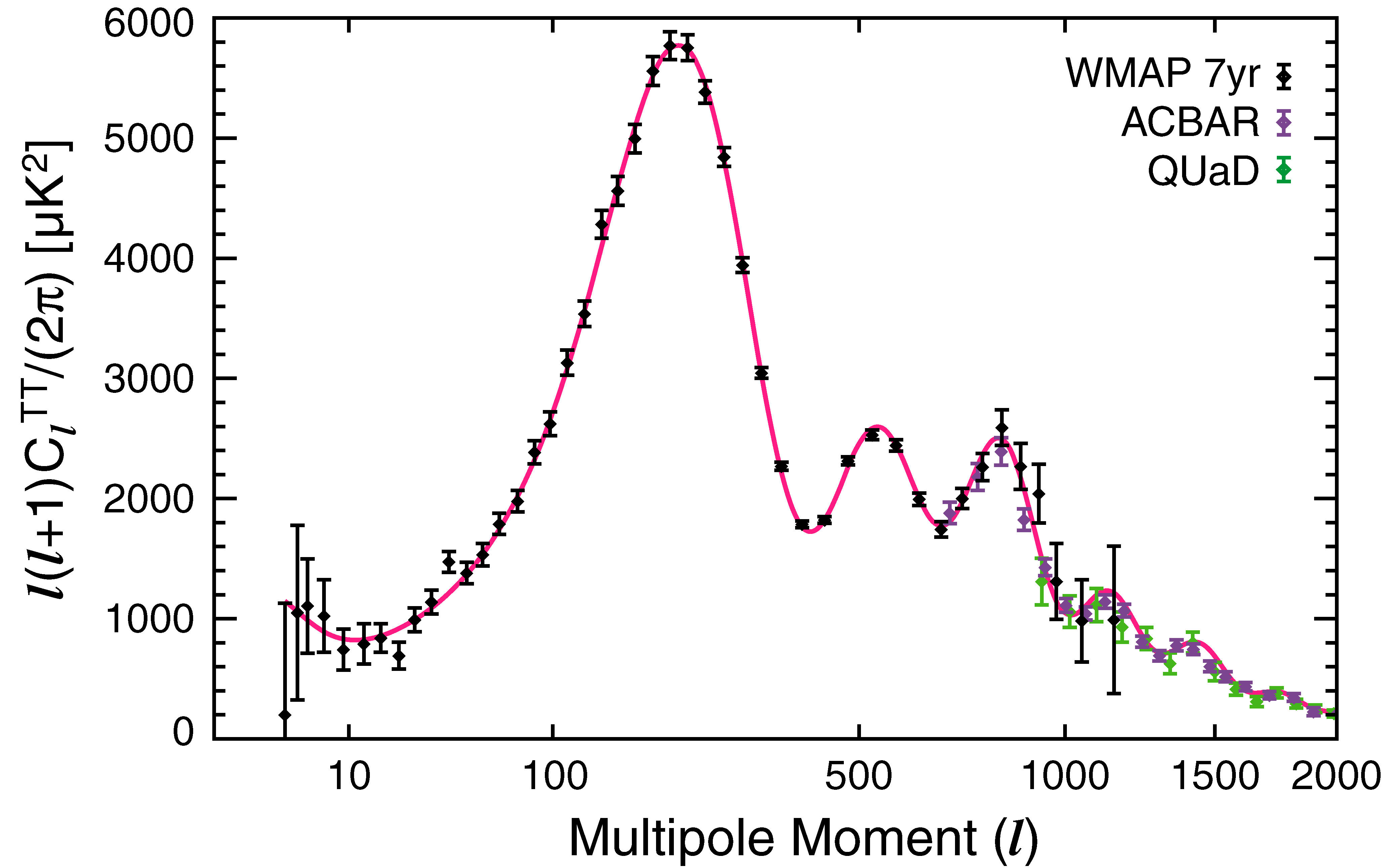}
%
%
%
\caption{A theoretical fit to CMBR data from the WMAP, ACBAR and QUAD experiments \cite{WMAP_komatsu}.
($C_l^{TT}$ is $C_l$ in the text.)}
\label{fig:WMAP7}
\end{figure}

From observations of the CMBR temperature and polarisation one obtains the $C_l$'s associated
with the 2 point temperature-temperature correlation $\langle T T\rangle$, 
polarisation-polarisation correlation
$\langle E E\rangle$, and temperature-polarisation correlation $\langle T E\rangle$, where $T$ and
$E$ refer
to the temperature and $E$ mode polarisation of the CMBR. 
One then fits the theoretical $C_l$'s to the $C_l$'s from observations by varying the 
parameters of the power spectrum 
$[A, n, dn/d(\ln k), r, n_T$  at some scales $k_{0,*}$] 
and cosmological parameters $[H_0, \Omega_{\rm m}, \Omega_{\rm de}, N_{\rm eff},..]$ and thereby obtains the
allowed ranges or bounds of all these parameters. ($H_0$ is the Hubble parameter today, $\Omega_{\rm m,de}$ 
represent the matter and dark energy density today, and $N_{\rm eff}$ is the effective number of relativistic neutrinos.)  Fig. \ref{fig:WMAP7} shows data from different experiments (WMAP, ACBAR and QUAD) and
a theoretical fit to the data.

\begin{figure}
\sidecaption
\includegraphics[width=7cm]{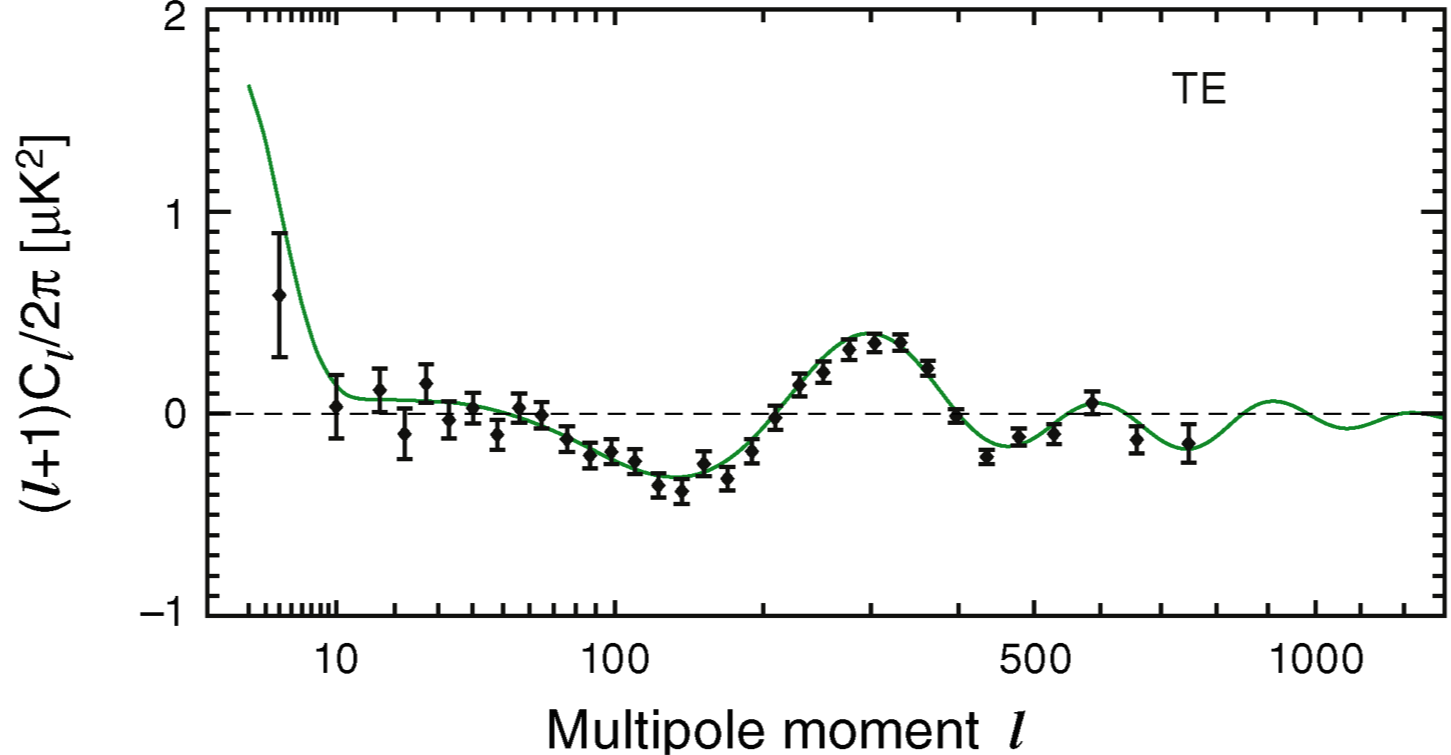}
\caption{
Anticorrelation between the temperature $T$ and $E$ mode polarisation 
between $l$ of 50-200 or $5^\circ > \theta >1^\circ$ \cite{WMAP_jarosik}, indicating superhorizon perturbations at decoupling.
}
\label{fig:TE}
\end{figure}

WMAP, the next satellite-based experiment after COBE, was launched in 2001
and collected 9 years of
data.
The WMAP data indicated that
\begin{itemize}
\item
Cosmological curvature perturbations are nearly
scale invariant, i.e. in the power spectrum $P(k)\sim A k^{n-1}$, $n\sim 1$.  

\item
There is an anticorrelation between $T$ and $E$ 
in $C_l^{TE}$ seen between $l$ of 50-200 or $5^\circ > \theta >1^\circ$ (see Fig. \ref{fig:TE}),
implying superhorizon perturbations at the time of decoupling, as predicted
by inflation \cite{Dodelson2003}.  
(Non-zero $C_l$  
corresponding to separation
angles greater than $1^\circ$ can be because of ISW, or other causal processes \cite{SpergelZald,Turok1,Turok2}.)

\item
There are well-defined peaks in the CMBR fluctuations -- these are consistent with inflation (but not with other mechanisms of producing primordial fluctuations, such as cosmic strings).

\end{itemize}
Furthermore, WMAP data
hinted at the possibility of large $f_{\rm NL}\sim100$ which suggested the possibility of multifield inflation, inflation with non-standard kinetic terms, etc.
Note that 
the skewness  
was still smaller than 1, i.e. the temperature fluctuations are largely Gaussian \cite{baumann}.

\begin{figure}
\sidecaption[t]
\includegraphics[width=7cm]{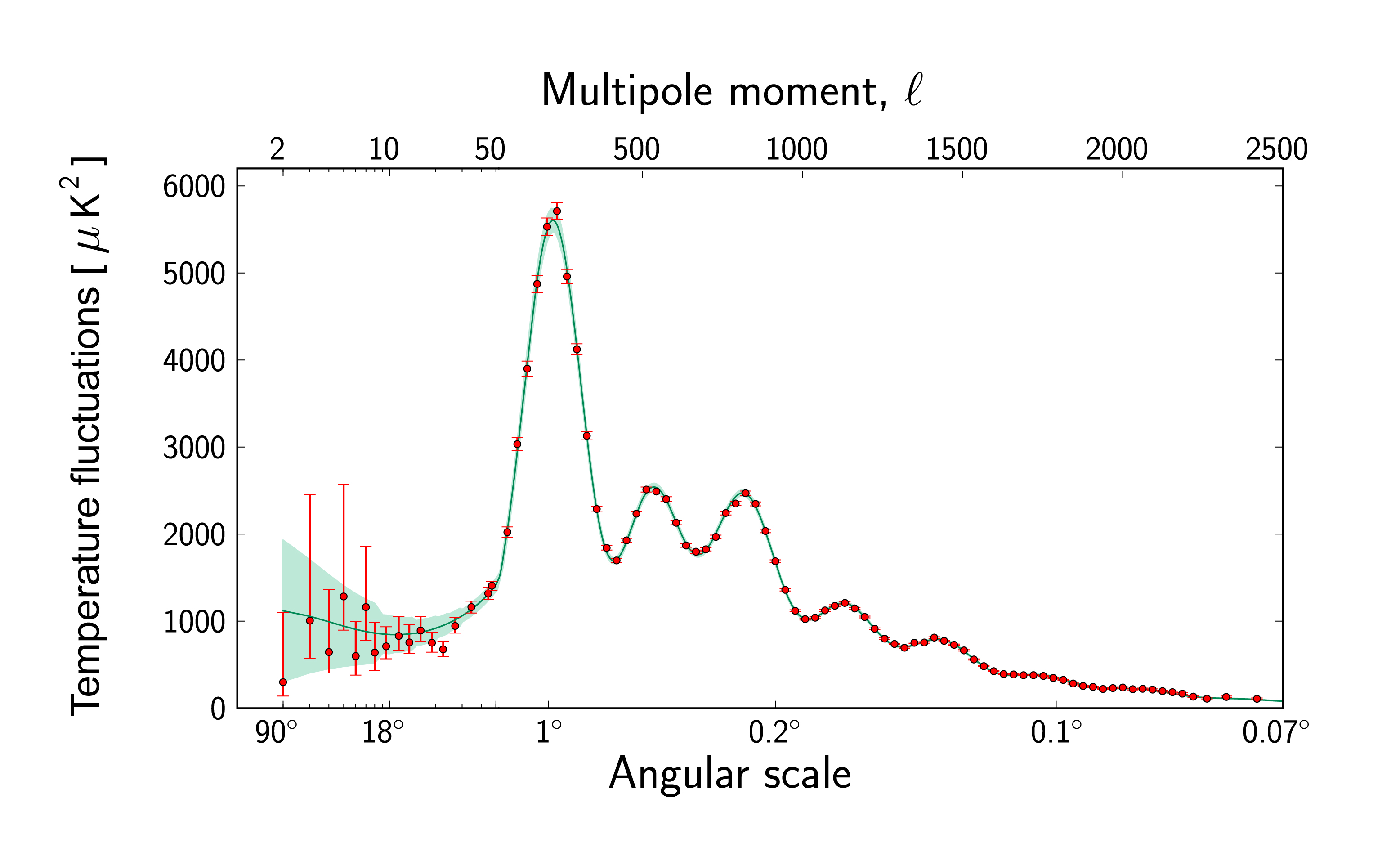}
\caption{
$C_l$ as measured by Planck (with a theoretical fit to the data). [Courtesy: ESA]
}
\label{fig:PlanckCl}
\end{figure}

The data from
Planck (2009-2013) lowered the error bars on the $C_l$'s and went to much smaller angles than WMAP (see Fig. \ref{fig:PlanckCl}).  
The Planck data, along with
other data, implied 
$n=0.9603\pm0.0073$ \cite{Planck2013CosmoParam}.
There was also a change in the mean values of cosmological parameters $H_0,  \Omega_{\rm m}$ and $\Omega_{\rm de}$ 
from the WMAP implied values but both sets of values were within 2 sigma of each other.
Furthermore, Planck data implied that $f_{\rm NL}$ was consistent with 0, and gave an upper bound on the tensor to scalar ratio, $r < 0.11$. This allowed us to rule in and rule out inflation models better than with WMAP data.

\subsection{Relating CMBR Observables to Inflation Models}
\label{subsec:2}
\begin{figure}
\sidecaption
\includegraphics[width=7cm]{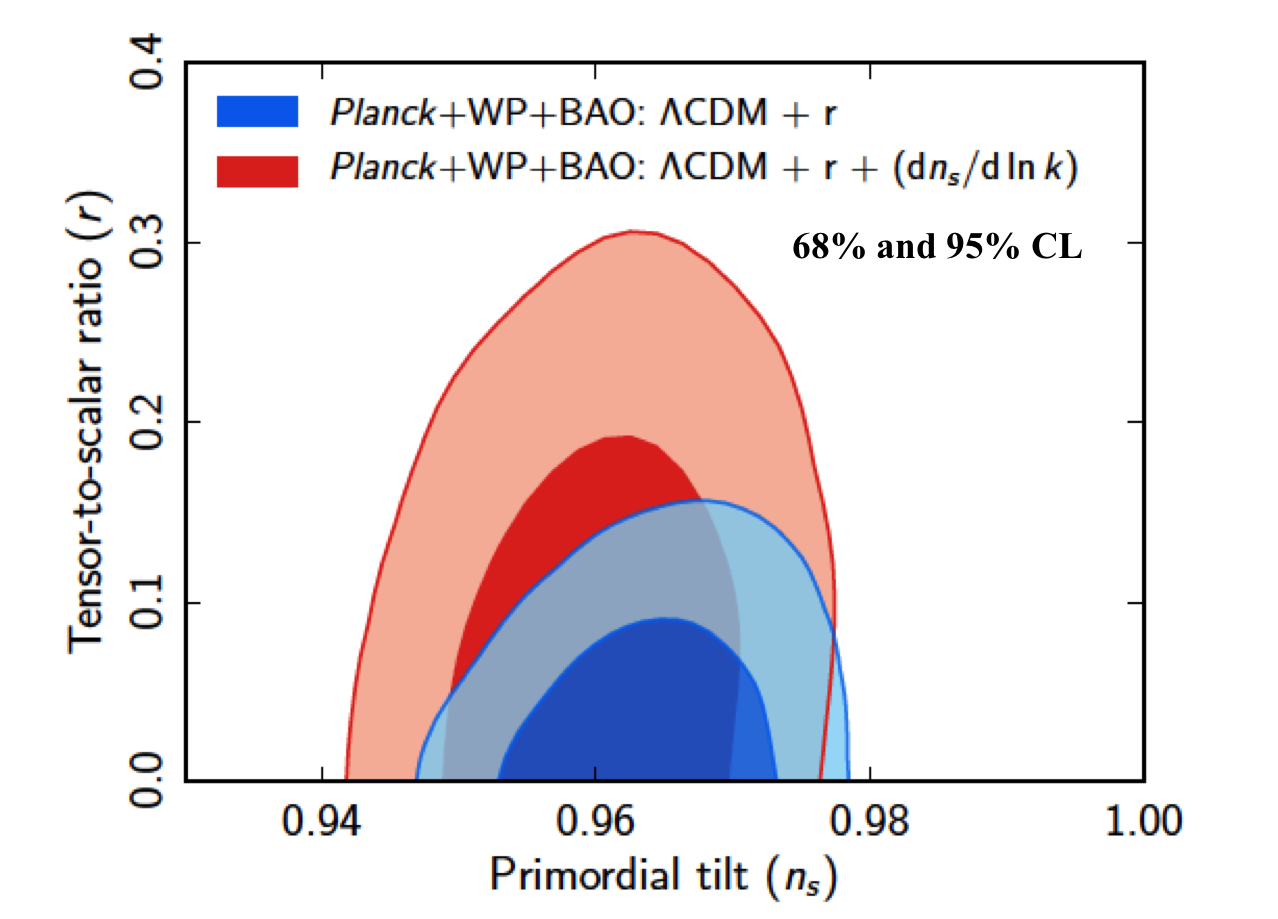}
\caption{
Likelihood contours in the $n-r$ plane after marginalising over other parameters, using data from Planck, polarisation
data from WMAP, and large scale structure data for baryon acoustic oscillations (BAO), for a $\Lambda$CDM model with
$r$, and $r$ and $dn/d(\ln k)$ as additional parameters.  ($n_s$ in this plot corresponds to $n$.) [From Ref. \cite{Planck2013Infl}.]
}
\label{fig:nrplot}
\end{figure}

As mentioned earlier, one can fit the observational data from CMBR, and from large scale structure, with 
theoretical realisations for a range of values for the parameters of the power spectra, and other cosmological
parameters.
The analysis can then be presented as likelihood contour plots in, for example, the $n-r$ plane after marginalising
over all other parameters, as in Fig. \ref{fig:nrplot}. 
The likelihood plots can give the allowed values of parameters, as mentioned
above.
Moreover, different inflation models can be projected onto such plots, and can 
then be ruled in or out.  

Inflaton models can be parametrised by $V(\bar\varphi) $, the height of the inflaton potential at some 
field value $\bar\varphi$ 
and slow roll parameters that depend on derivatives of the inflaton potential.  The first few slow roll parameters are given by
\begin{equation}
\epsilon=\frac{\mpl^2}{16\pi}\left(\frac{V^\prime(\varphi)}{V(\varphi)}\right)^2
\qquad
\eta=\frac{\mpl^2}{8\pi}\left(\frac{V^{\prime\prime}(\varphi)}{V(\varphi)}\right)
\qquad
\xi=\frac{\mpl^2}{8\pi}\left(\frac{V^\prime V^{\prime\prime\prime}}{V^2}\right)^\frac12\,,
\end{equation}
where slow roll implies a flat potential, and so $\epsilon, \eta$ and $\xi\ll 1$.
Then
\begin{eqnarray}
n = 1 - 6 \epsilon + 2 \eta \qquad &     r = 16 \epsilon    \nonumber\\
	dn/d(\ln k) = 16 \epsilon\eta - 24 \epsilon^2 - 2 \xi^2\qquad
&
	V = (1.9 \times 10^{16} \gev)^4 (r/0.12)\,.
	\end{eqnarray}
	For each inflation model, one can obtain possible values of $n, r,$ etc. from the parameters of the model,
	and compare the former with the allowed ranges from the data to 
assess the viability of the model.  (An alternate Bayesian approach can be found
in Ref. \cite{martin2014}.)

We remind the reader here of the time scales relevant to the generation and observation of CMBR anisotropies.  
The scalar and tensor perturbations represented by the power spectra $P(k)$ and $P_T(k)$ are generated during
inflation at some time between $10^{-38}$s to $10^{-11}$s (for GUT scale to electroweak scale inflation), which
imprint fluctuations in the temperature and polarisation of the CMBR at $t_{\rm dec}=380,000$ yr, which manifest themselves as 
temperature and polarisation anisotropies as observed by us today at $t=13.8$ billion years.

\section{Models of Inflation and Planck}

$f_{\rm NL}$ consistent with zero and lack of isocurvature perturbations, as seen by Planck, favours simpler single-field slow roll inflation models with standard kinetic terms.  Some of these models are new inflation or hilltop inflation models with concave or Coleman-Weinberg type potentials, chaotic inflation models with monomial potentials like
$V(\varphi) \sim m^2\varphi^2, \lambda\varphi^4$, natural inflation models with a pseudo-Nambu-Goldstone boson as the
inflaton, hybrid inflation models popular in the context of SUGRA as D and F term inflation, and the Starobinsky model involving
a modification of Einsteinian gravity.  One may also consider non-minimal inflation models where the Lagrangian includes
a non-minimal coupling of the inflaton to gravity with a term in the Lagrangian as $1/2 \,\xi R \varphi^2$, where $\xi$ here
is a coupling constant (not to be confused with the slow roll parameter above) and $R$ is the Ricci scalar (not the scale factor).  Higgs inflation,
in which the Higgs field plays the role of the inflaton is an example of non-minimal inflation with $\xi\sim10^5$.  There are 
many more models of inflation and Ref. \cite{martin_encycl} discusses 70 more such models.

\begin{figure}
\sidecaption
\includegraphics[width=7cm]{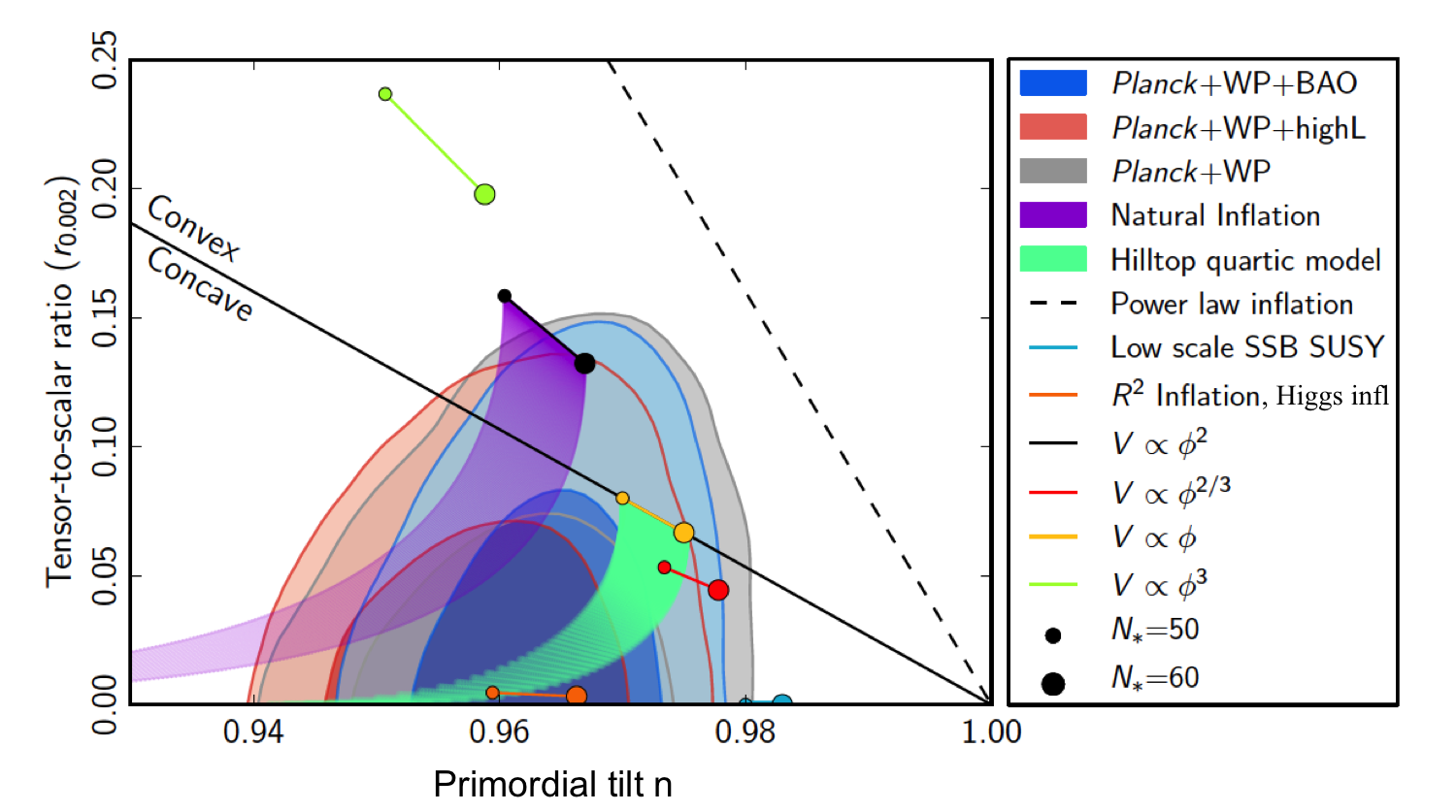}
\caption{
Likelihood contours in the $n-r$ plane, as in Fig. \ref{fig:nrplot}, with various inflation models projected on to the plane for 50-60 e-foldings of 
inflation (reflecting our uncertainty in the details of reheating).  High $l$ refers to data from the Atacama Cosmology Telescope (ACT) and
South Pole Telescope (SPT).
[From Ref. \cite{Planck2013Infl}.]
}
\label{fig:PlanckInflationModels}
\end{figure}

Projecting these models onto the $n-r$ allowed contour plane from Planck data
indicates that hilltop, natural, Starobinsky and Higgs inflation
models are preferred while
exponential potential models (giving power law inflation with $R(t)\sim t^q, q>1$),  simplest hybrid models, chaotic inflation models 
with the exponent in the potential $p>2$ do 
not fit the data well, as seen in Fig. \ref{fig:PlanckInflationModels}.
For non-minimal inflation, the quartic chaotic inflation model with coupling $\lambda\sim10^{-14}$ is viable for 
$\xi>10^{-3}$ \cite{bezrukov_gorbunov,okada_etal2010}. Furthermore, the bound
$r<0.11$ implies $V^{1/4} < 1.9 \times 10^{16}\gev$.

\section{BICEP2}

BICEP (Background Imaging of Cosmic Extragalactic Polarization) is a
ground based (South Pole) experiment.  It claimed to have
detected evidence of  $B-$mode polarisation in the CMBR.
These can only be generated by tensor perturbations of the metric, or gravitational waves.
Evidence of tensor modes on superhorizon scales at decoupling is a smoking gun for inflation.  
BICEP2 reported they had detected 
$r\sim 0.2$ which translated to $V^{1/4} = 2 \times 10^{16} \gev$.        
But the BICEP2 result was in conflict with $r < 0.11$ of Planck (from contributions to 
$T-T$ correlations).  Note that measurements by these
two experiments are on different scales.  There have been attempts to explain this discrepancy by
considering running of the scalar spectral index (i.e. $dn/d(\ln k)\ne0$), or $n_T>0$ which are in conflict with 
certain observations \cite{eastherpeiris,gerbino_etal2014,wu_etal2014}.  However increasing $N_{\rm eff}$ may help resolve the tension \cite{Giusarma_etal2014,Zhang_etal2014,Dvorkin_etal2014,Li_etal2014}. 

It has been argued that the polarisation detected by BICEP2 may be from light scattering off dust in our galaxy rather than having a primordial origin \cite{MortonsonSeljak2014,FlaugerHillSpergel2014}.
But Ref. \cite{ColleyGott2014} has argued that there is an
imperfect match between Q and U Stokes parameter maps for Planck and BICEP2  which implies that half the 
BICEP2 signal can not be due to dust,
and obtained $r=0.11\pm0.04$.		    
The Planck and BICEP2 collaborations
have also worked together to resolve the discrepancy and in 
Ref. \cite{PlanckBICEP2} have stated
that there is no statistically significant evidence for tensor metric perturbations.


\section{Additional Issues}

Below we list some additional issues related to inflationary cosmology, or its
alternatives.

\subsection{Low Power on Large Scales}

The observed angular power spectrum by Planck (the red dots in Fig. \ref{fig:PlanckCl}), for $l<30$, 
is lower than what would be expected
from the standard $k^{n-1},\,\, n\sim 1$ scalar power spectrum that one predicts for inflation (green line
in Fig. \ref{fig:PlanckCl}).   This indicates that the scalar power spectrum has less power on large scales.  The 
decreased power in the angular power spectrum at large angles
was observed by COBE and WMAP too.  Cosmic variance associated with lesser sampling possibilities  for large 
angles gives large uncertainties at these angles (the theoretical uncertainty is included in the error bars in 
Fig. \ref{fig:PlanckCl}).  If inflation lasts only the required number of e-foldings to solve the horizon and flatness problems
and is preceded by a radiation dominated era, it would provide a natural cutoff to the scalar power spectrum on large
scales which could explain the lower power in the angular power spectrum at low $l$.  Note that the ISW effect should also
give some contribution to the angular power spectrum at low $l$, as indicated in Fig. \ref{TheoreticalClPlot}.

\subsection{Homogeneous Initial Conditions}

For inflation to start the inflaton field needs to be uniform on
scales somewhat larger than the causal horizon at the beginning of inflation 
\cite{FarhiGuth,GoldwirthPiran,KungBrandenberger, 
TroddenVachaspati}.
Invoking the string landscape, multiple stages of inflation and the anthropic principle,
one may argue that the Universe tunnelled to the current vacuum through bubble
nucleation and that the Universe was homogeneous on the required
scale at the beginning of the final stage of inflation which is the stage relevant for
observations \cite{GuthKaiserNomura2014}.

\subsection{The eta Problem}

The slow roll parameter, $\eta$, is (for a quadratic inflaton potential) 
\begin{equation}
\eta=
\frac{m^2}{3H^2}
\,.
\end{equation}
One expects quantum corrections to take $m$ to $M_{\rm Pl}\gg
H$.  Alternatively, in a supersymmetric theory, the corrections should have a cutoff at  the Hubble
parameter $H$, since the finite energy density of the inflaton breaks supersymmetry, and so $m\sim H$.  In either case, $\eta$ will be too large.
One may invoke a shift symmetry $\varphi\rightarrow \varphi+ {\rm constant}$ to protect the 
inflaton mass, but note that quantum gravity effects can break global symmetries.

\subsection{Trans-Planckian Problem}

Density fluctuations that we see today are due to quantum fluctuations during inflation.  For cosmologically
relevant length scales,
the corresponding modes in the power spectrum would have a physical wavelength smaller than the Planck length at the onset of inflation (when we impose initial conditions). This is referred to as the Trans-Planckian problem.
This indicates that the predictions of inflation may not be robust given
uncertainties related to physics beyond the Planck scale.  One then  
includes modifications in the scalar power spectrum based on some assumptions
or understanding of
 Planck scale physics.
Attempts of the latter include a			
modification of the  initial conditions of the fluctuation modes, imposing initial conditions on modes once they cross the
Planckian threshhold, etc. \cite{BrandenbergerMartinReview}.

\subsection{Breakdown of Statistical Isotropy}

An analysis of the CMBR temperature anisotropy indicates
a preferred axis, i.e., an orientation of the dipole, quadrupole and octopole components along a particular axis.
Radio polarisation data from radio galaxies
and optical polarisation data from quasars also lead to a similar conclusion.
All these signals of a breakdown of statistical isotropy are associated with an axis 
directed towards the Virgo cluster \cite{RalstonJain2004}.  Studies are on to determine whether the signal
is real or due to foreground. 

\subsection{Higgs Instability and Inflation}

Theoretical calculations indicate that for large values of the Higgs field, the potential is
unstable (depending on the top quark mass).
If so, quantum fluctuations or tunneling during inflation can
cause the Higgs field to be in the unstable region
\cite{KobakhidzeSpencerSmith,EnqvistMeriniemiNurmi}.  The associated
negative energy density can then overcome the
positive energy density driving inflation
and the Universe can collapse into a black hole.  It has also been argued that if one considers the Hawking-Gibbons temperature $H_I/(2\pi)$ during inflation then the effective potential is stable and collapse can be avoided \cite{GoswamiMohanty2014}.

\subsection{Alternatives to Inflation}

We mention here that an ekpyrotic or cyclic Universe has been considered as an
alternative to inflation \cite{SteinhardtVaasreview}.  In such a scenario, which involves higher dimensions, 
our Universe lies on a brane
and goes through collapsing and expanding
phases because of collisions with another brane.
This scenario
can explain the horizon problem as regions come in
causal contact during collapse, and the flatness problem.
Scalar perturbations are also generated with superhorizon
correlations because of quantum fluctuations on branes
before collisions. Thus it reproduces many of the positive features of 
inflationary cosmology.  Note that in this scenario
primordial tensor perturbations are small, and $n_T >0$, and the
tensor perturbations are induced by scalar
perturbations \cite{BaumannSteinhardtTakahashiIchiki}.


\section{Conclusion}

To summarise, the inflationary paradigm is well consistent with data.  Observations
of the temperature anisotropies indicate that the scalar perturbations are 
nearly scale invariant, superhorizon, and primarily Gaussian.  The scale 
of inflation is less than or equal to $10^{16} \gev$.  The CMBR data has become
increasingly more precise and it is now possible to rule out many specific models 
of inflation.  The joint efforts of the BICEP and Planck collaborations has indicated
that the earlier BICEP2 result of a relatively large tensor-to-scalar ratio has to be revised
to an upper bound on $r$ at a lower value, consistent with Planck's earlier results.
Future experiments, e.g. BICEP3, will hopefully shed more
light on aspects of the inflationary epoch in the early Universe.

\end{document}